\documentclass[aps,superscriptaddress,nofootinbib,
prd,notitlepage,twocolumn,showkeys]{revtex4-1} 

\pdfoutput=1

\usepackage{amsfonts}
\usepackage{amsmath}
\usepackage{amssymb}
\usepackage{graphicx,color}
\usepackage{float}
\usepackage{hyperref}
\usepackage{subfigure}
\usepackage{dcolumn}% Align table columns on decimal point
\usepackage{soul}
\usepackage{ulem}
\usepackage{verbatim}

\begin{document}

%---------------------------------------------------------------------%
\title{Resurrecting Vanilla Power Law Inflation with the aid of \\ Continuous Spontaneous Localization in the {\it ACT} era}
%---------------------------------------------------------------------%

\author{Suratna Das}
\email{suratna.das@ashoka.edu.in}
\affiliation{Department of Physics, Ashoka University,
   Rajiv Gandhi Education City, Rai, Sonipat: 131029, Haryana, India}

%%%%%%%%%%%%%%%%%%%%%%%%%%%%%%%%%%%%%%%%%%%%
\begin{abstract}

The Vanilla Power Law Inflation is plagued with two severe drawbacks, the one being the issue of graceful exit, and the other being its compatibility with the existing data. There's yet another daunting problem generic to any inflationary model, and not particular to the Power Law Inflation, is the issue of classicalization of the primordial quantum perturbation. This issue is often treated as an isolated problem primarily because decoherence is often invoked to tackle such problems and decoherence, in principle, doesn't leaves any observational imprints. However, that's not the case if one invokes the collapse models of quantum mechanics to resolve such a problem, because the collapse models  do modify the Schr\"{o}dinger evolution of the quantum system and the modified dynamics is bound to leave imprints. We show that collapsed modified Power Law Inflation can indeed circumvent the issue with observations and can be made compatible with all the current data coming from {\it Planck}, {\it ACT}, DESI and BAO while resolving the classicalization issue associated with the quantum primordial fluctuations. Such an inflationary model also produces a more red-tilted tensor spectrum and no running for both the scalar and tensor spectral indices, which can be a litmus test for this model. 

\end{abstract}
%%%%%%%%%%%%%%%%%%%%%%%%%%%%%%%%%%%%%%%%%%%

\maketitle

%%%%%%%%%%%%%%%%%%%%%%%%%%%%%%%%%%%%%%%%%%

{\it Implications of the recent ACT results:} The recently published Atacama Cosmology Telescope ({\it ACT}) data \cite{ACT:2025fju, ACT:2025tim} are suggesting a less red-tilted scalar power spectrum than the one measured by {\it Planck} by pushing the value of the scalar spectral index $n_s$ more towards unity. The combined analysis of {\it Planck} \cite{Planck:2018vyg} with {\it ACT}  \cite{ACT:2025fju, ACT:2025tim} (including CMB lensing) along with the BAO data from the 1st year DESI results \cite{DESI:2024uvr, DESI:2024mwx} (which we will refer to as P-ACT-LB henceforth) sets the value of the scalar spectral index as 
\begin{eqnarray}
n_s=0.9743\pm0.0034.
\label{ns-act}
\end{eqnarray}
This measured value of $n_s$ differs by $2\sigma$ from the previous {\it Planck} measurements ($n_s=0.9651\pm0.0044$) \cite{Planck:2018vyg}. 

If this result stands the test of time, then the {\it Planck}-preferred inflationary models, e.g. Starobinsky model \cite{Starobinsky:1980te}, Higgs inflation model \cite{Bezrukov:2007ep} and the $\alpha$-attractor models \cite{Kallosh:2013yoa} that used to sit in the sweet-spot of the $n_s-r$ plot ($r$ being the tensor-to-scalar ratio) \cite{Planck:2018vyg}, will all be disfavoured by $2\sigma$. This opens up the window to re-evaluate the models which were previously discarded for not making a space in the {\it Planck} $n_s-r$ plot and/or to come up with new inflationary models with less red-tilted spectrum. These inflationary models not only require to yield the new value of $n_s$ from the  joint measurements of the  P-ACT-LB, but also have to comply with the latest bounds on $r$ set by the joint analysis of {\it Planck} and BICEP as  \cite{BICEP:2021xfz}
\begin{eqnarray}
r<0.036.
\label{bound-r}
\end{eqnarray}
A few attempts have already been made in the literature in this direction. There has been attempts to revive the monomial chaotic potentials (through non-minimal coupling \cite{Kallosh:2025rni} or via warm inflation \cite{Berera:2025vsu}) as well as the Higgs Inflation \cite{Aoki:2025wld, Gialamas:2025kef}. A new model has been proposed \cite{Salvio:2025izr} where inflation is driven by a dynamical affine connection which is compatible with these observations. 

In this letter, we propose a mechanism of reviving the vanilla power law inflationary model \cite{Lucchin:1984yf} in the light of the recent P-ACT-LB data. 

{\it Vanilla power law inflation and its known problems:} It was pointed out in 1985 in \cite{Lucchin:1984yf} that instead of having an exponential expansion during an inflationary phase, which is a mere particular case, one can incorporate a power law expansion ($a(t)\propto t^q$, $q\gg1$) during inflation while addressing the fine-tuning problems of the Big Bang paradigm, e.g. horizon problem and flatness problem, with ease. Such an inflationary phase was dubbed Power Law Inflation (PLI), and it was noted that such a power law expansion is driven by a particular form of the inflaton potential given as 
\begin{eqnarray}
V(\phi)=V_0\exp\left(-\lambda\phi/M_{\rm Pl}\right),
\end{eqnarray}
where $\lambda\equiv\sqrt{2/q}$ is a constant and $M_{\rm Pl}$ is the reduced Planck mass. A signature of the PLI is that it yields constant slow-roll parameters:
\begin{eqnarray}
\epsilon_V=\frac{\lambda^2}{2},\quad\quad \eta_V=\lambda^2,
\end{eqnarray}
which then gives rise to $n_s$ and $r$ as follows:
\begin{eqnarray}
n_s-1&=&2\eta_V-6\epsilon_V=-\lambda^2,\nonumber \\
r&=&16\epsilon_V=8\lambda^2.
\label{ns-r}
\end{eqnarray}

Though quite appealing, this vanilla PLI has a couple of well-known serious drawbacks:
\begin{enumerate}
\item The constant slow-roll parameters straightforwardly indicate that PLI possess graceful exit problem. 
\item The {\it Planck} observations \cite{Planck:2018vyg} restricts the scalar spectral index in the range $[0.9607\,-\,0.9695]$ at 95\% CL, which then yields $\lambda$ to lie within the range $[0.175\,-\,0.198]$ ($51\lesssim q\lesssim 65$). These bounds, on the other hand, yields $r$ as $0.24<r<0.31$ which is way too large in comparison to the upper limit on $r$ set by {\it Planck} and BICEP \cite{BICEP:2021xfz}. 
\end{enumerate}
These couple of points render the vanilla PLI model observationally disfavoured. 

However, there is one more long-standing issue, though not particular to PLI itself but to any generic inflationary model, and that is the question of classicalization of the primordial quantum perturbations. Decoherence has been proposed and considered extensively over the years as a mechanism to encounter such problems in inflationary dynamics \cite{Kiefer:2008ku, Kiefer:2006je, Kiefer:1998qe, Polarski:1995jg}. However, as decohence doesn't modify the dynamics of the inflaton, it doesn't leave any observational imprints which renders such mechanism observationally non-falsifiable. 

On the other hand, collapse models of quantum mechanics \cite{Bassi:2003gd}, in particular Continuous Spontaneous Localization (CSL) model, were proposed to encounter the quantum-to-classical transition problem in laboratory systems (for a more recent review see \cite{Bassi:2012bg}). The collapse models of quantum mechanics phenomenologically modify the Schr\"{o}dinger equation by adding non-linear and stochastic terms in order to address the quantum-to-classical transition problem in the laboratory systems. The collapse-modified dynamics is thus bound to leave observational imprints. CSL has been explored to encounter the classicalization problem of inflationary primordial quantum perturbations before \cite{Canate:2012nwv, Martin:2012pea, Das:2013qwa}, and there too, the modified dynamics leaves its imprints in the observables, like $n_s$ \cite{Das:2014ada}. 

We will now briefly discuss how CSL modifies the observables in any standard inflationary picture, and then we will show how such modified dynamics can help make vanilla PLI observationally viable.

{\it CSL-modified inflationary dynamics \& the observables:} A few approaches have been devised in the literature to incorporate CSL mechanism into inflationary dynamics  \cite{Canate:2012nwv, Martin:2012pea, Das:2013qwa}. We will follow the approach developed in \cite{Martin:2012pea, Das:2013qwa} as that is a direct adaptation of the CSL mechanism applicable to laboratory systems into inflationary dynamics. As CSL dynamics modifies the Schr\"{o}dinger equation of a quantum system, the inflationary perturbation theory needs to be analysed in the Schr\"{o}dinger picture in order to incorporate the CSL mechanism. According to the method prescribed in \cite{Martin:2012pea, Das:2013qwa}, the wave-functional of the primordial scalar perturbations  $\Psi[\zeta(\tau,{\mathbf x})]$ (where $\zeta$ is the Mukhanov-Sasaki variable) follows a CSL-modified functional Sch\"{o}dinger equation in the momentum space which can be written as 
\begin{eqnarray}
d\Psi_{\mathbf k}&=&\left[-i{\mathcal H}_{\mathbf k}d\tau+\sqrt{\gamma}\left(\zeta_{\mathbf k}-\langle\zeta_{\mathbf k}\rangle\right) dW_\tau\right.\nonumber\\
&&\left.\,-\frac\gamma2\left(\zeta_{\mathbf k}-\langle\zeta_{\mathbf k}\rangle\right)^2d\tau\right]\Psi_{\mathbf k},
\end{eqnarray}
where the first term on rhs represents the standard Schr\"{o}dinger equation (with ${\mathcal H}$ being the Hamiltonian) whereas the following two terms are added CSL terms. The second term on the rhs contains the Wiener process $W_\tau$ that turns the CSL-modified Schr\"{o}dinger equation to be a stochastic one, and the third term is the non-linear term which essentially restricts superposition of wavefunctions. Both these qualities, stocasticity and non-linearity, are beyond the features of the standard Schr\"{o}dinger equation, and their strength is controlled by the collapse parameter $\gamma$. Here, $\tau$ is the conformal time. 

It is well-known that the sub-horizon quantum perturbations behave quantum mechanically during inflation, whereas they turn more classical after crossing the horizon and when they become super-horizon. Thus the collapse parameter, which controls the strength of the collapse should distinguish between sub- and super-horizon modes by remaining negligible for the sub-horizon modes while grow as the modes tend to become super-horizon. Initially CSL with a constant $\gamma$ (i.e. the strength of collapse is same for all modes, be it sub-horizon or super-horizon) was analyzed in \cite{Martin:2012pea} and no successful classicalization of quantum primordial perturbations were obtained. Then in \cite{Das:2013qwa}, CSL was analyzed with the form of $\gamma$:
\begin{eqnarray}
\gamma=\frac{\gamma_0(k)}{(-k\tau)^\alpha}, \quad\quad 1<\alpha<2,
\end{eqnarray}
which is negligible for sub-horizon modes $-k\tau\rightarrow \infty$ and grows while the modes become super-horizon $-k\tau\rightarrow0$. Such parametrizations was shown to explain the classicalization of the quantum modes successfully. Finally, to obtain a scale-invariant spectrum, another parameter $\beta$ was introduced via  
\begin{eqnarray}
\gamma_0(k)=\tilde\gamma_0(k/k_0)^\beta,
\end{eqnarray}
that yields a scalar spectral index \cite{Das:2014ada}
\begin{eqnarray}
n_s-1=\delta+2\eta_V-6\epsilon_V,
\label{ns-csl}
\end{eqnarray}
where $\delta=3+\alpha-\beta$. 

It is then natural to assume that such collapse mechanism will also be effective for the tensor modes \cite{Das:2014ada}, and that will then yield a tensor spectral tilt:
\begin{eqnarray}
n_T=\delta-2\epsilon_V,
\label{nT-csl}
\end{eqnarray}
while yielding the same tensor-to-scalar ratio 
\begin{eqnarray}
r=16\epsilon_V.
\end{eqnarray}
It is to note that, these analysis are all phenomenological and a quantum field theory of CSL mechanism is yet to be devised. If and when such a quantum field theory of CSL is developed, it may turn out that $3+\alpha=\beta$ yielding $\delta$ to be identically zero (which should arise from a deeper reason than just being a mere coincidence). However, if that doesn't happen, then we can see that CSL modified inflationary dynamics indeed leave observational imprints by modifying the standard spectral tilts, both of the scalar and the tensor spectra. And, as $\delta$ appears in the spectral tilt, its value can at best be of the order of slow-roll parameters ($\mathcal{O}(10^{-2})$) and can assume either positive or negative values. Any field theoretic version of CSL yielding a larger value of $\delta$ will then be ruled out by cosmological observations. Therefore, it seems that cosmology has already set some guidelines for any field theoretic version of CSL that are yet to be developed. 

We will exploit these modifications to the spectral tilts due to the incorporated CSL dynamics to revive the vanilla PLI. 

{\it CSL-modified PLI:}  As the CSL-modified dynamics leave the sub-horizon evolution of quantum modes untouched, it is then straightforward to predict $n_s$ and $r$ in the CSL-modified PLI:
\begin{eqnarray}
n_s-1&=&\delta-\lambda^2,\nonumber\\
r&=&8\lambda^2,
\end{eqnarray}
as opposed to the ones given in Eq.~(\ref{ns-r}). Therefore, to respect the current observational bound on r (as given in Eq.~(\ref{bound-r})), one requires $\lambda<0.067$ ($q>444$). This upper bound on $\lambda$ along with the upper bound on $n_s$ from the P-ACT-LB joint analysis (as given in Eq.~(\ref{ns-act})) puts an upper bound on the CSL parameter $\delta$ as $\delta<-0.0178$. With these bounds on $\lambda$ and $\delta$, the two model parameters of CSL-modified PLI, this inflationary scenario is quite in accordance with the current measurements of P-ACT-LB. 

It is important to note that the CSL modified PLI will have two distinctive features apart from being in accordance with the observations:
\begin{enumerate}
\item The tilt of the tensor spectrum in CSL-modified PLI will be $n_T=\delta-2\epsilon_V$ \cite{Das:2014ada}. In such  case, as $\delta$ needs to be negative for the scenario to be in tune with the observations, the tensor spectrum will be more red-tilted than the standard scenario. 
\item As $n_s$ and $n_T$ are both constants in the CSL-modified PLI, the running of both these indices will be exactly zero. 
\end{enumerate}
These two features can be used as a litmus test of the CSL modified PLI scenario once the primordial tensor spectrum is observationally detected and measured. 

{\it In conclusion,} the CSL-modified PLI can not only explain successfully the classicalization of the primordial quantum perturbations (both scalar and tensor) but also be in tune with all the current cosmological observations, i.e. {\it Planck}, {\it ACT}, DESI and BICEP. The more red tilted tensor spectrum than a standard inflationary scenario as well as no running of both the scalar and tensor spectral tilts are smoking gun signatures of this inflationary scenario. However, like in the standard PLI, one needs to invoke some mechanism to gracefully exit from the inflationary phase, like by modifying the potential to $\cosh{\lambda\phi/M_{\rm Pl}}$ \cite{Sahni:1999qe} (which approximates to the exponential potential in large values of $\phi$) or by a perturbed modification to the power-law definition as suggested in \cite{Alhallak:2022szt} or by some waterfall mechanism \cite{Linde:1993cn, Copeland:1994vg}. 

It is interesting to note that to make CSL modified PLI work with the current cosmological observations one needs to have a negative CSL $\delta$ parameter. If CSL is applied to other viable inflationary models in order to answer the quantum-to-classical transition of the primordial perturbations, CSL will modify the observables $n_s$ and $n_T$ in those models too. It is evident from Eq.~(\ref{ns-csl}) and Eq.~(\ref{nT-csl}) that CSL shifts the spectral indices, both the scalar and the tensor one, by a constant amount $\delta$. Hence, one can write 
\begin{eqnarray}
n_s^{\rm CSL}=\delta+n_s,\quad\quad\quad n_T^{\rm CSL}=\delta+n_T,
\end{eqnarray}
where $n_s$ and $n_T$ are the spectral tilts of the inflationary model without CSL modifications. Moreover, the CSL dynamics will not alter the running of the spectral indices as the modification to the spectral indices is through a constant parameter $\delta$. There are other viable inflationary models which are preferred by {\it Planck} observations, like Starobinsky  inflation \cite{Starobinsky:1980te}, Higgs inflation \cite{Bezrukov:2007ep} and the $\alpha$-attractor models \cite{Kallosh:2013yoa, Kallosh:2013hoa, Kallosh:2022feu} that produce $n_s$ and $r$ that are in tune with {\it Planck}'s observations. Such models are now disfavoured by the recent ACT joint analysis \cite{Kallosh:2025ijd}. If CSL dynamics is applied in such inflationary models, then a negative $\delta$ of the order 0.01, required for making PLI viable with the observations, would shift these models from the sweet spot of the {\it Planck}'s observations, and it would shift it in the opposite direction which could have made these models viable for the recent P-ACT-LB joint analysis. On the other hand, if any future field theoretic construction of CSL suggests a positive $\delta$ of order 0.01, then that will shift the scalar spectral index of these models (Starobinsky inflation, Higgs inflation and $\alpha$-attractor models) in the right direction making them in tune with the current observations. Such models would then yield less red-tilted or even blue-tilted tensor spectrum depending on the value of $\delta$. In such a scenario, PLI will no longer be a observationally viable inflationary model. Therefore, it depends on future constructions of concrete  field theoretic version(s) of CSL to see which CSL modified inflationary dynamics would stand the test of time. To emphasise, the detection of tensor tilt as well as the runnings of the spectral tits are of utmost importance in order to pin down the correct inflationary model. 

%%%%%%%%%%%%%%%%%%%%%%%%%%%%%%%%%%%%%%%%%%%%%%%%%%%
\acknowledgements

The author would like to thank Varun Sahni and Swagat Mishra for suggesting some of the references. The author would like to thank the unknown referee as well for making very useful comments on the manuscript. The work of the author is supported by the Start-up Research Grant (SRG) awarded by Anusandhan National Research Foundation (ANRF), Department of Science and Technology, Government of India [File No. SRG/2023/000101/PMS].

%%%%%%%%%%%%%%%%%%%%%%%%%%%%%%%%%%%%%%%%%%%%%%%%%%%
\label{Bibliography}
\bibliography{pli-csl}

%%%%%%%%%%%%%%%%%%%%%%%%%%%%%%%%%%%%%%%%%%%%%%%%%%%

%\end{thebibliography}

\end{document}